\documentclass[twocolumn,showpacs,preprintnumbers,amsmath,amssymb]{revtex4}
\usepackage[dvips]{graphicx}
\usepackage{dcolumn}

\begin{document}

\title
{Effect of Dipole Energy on Half-Quantum Vortex in Superfluid Polar Phase}

\author{Tomohiro Hisamitsu and Ryusuke Ikeda}

\affiliation{%
Department of Physics, Kyoto University, Kyoto 606-8502, Japan
}

\date{\today}

\begin{abstract}
NMR experiments on superfluid $^3$He in a nematic aerogel have shown that the stability of the half-quantum vortex (HQV) realized in the polar phase depends on the relative direction between the magnetic field and the anisotropy axis brought by the aerogel. The vortex energy in the polar phase is examined in terms of the Ginzburg-Landau free energy incorporating the dipole energy, and the reason why the HQV in the polar phase in nematic aerogels became unstable in a transverse magnetic field  upon a field-cooling is explained. 

\end{abstract}



\maketitle

\section{Inroduction}
\label{sec:intro}
A novel high temperature superfluid phase has been discovered in liquid $^3$He in nematic aerogels. This equal-spin paired phase was identified with the polar pairing state through several observations \cite{Dmitriev15,Autti1,Autti2,Parpia}. Among them, one conclusive observation is the discovery \cite{Autti1} of the half-quantum vortex (HQV) in the polar phase. In the polar phase with a frozen orbital component of the order parameter, possible vortices are limited to the HQV and the simple phase vortex (PV) with no texture of the spin component ($d$-vector) accompanied. Hence, by applying a magnetic field, one can experimentally judge which of the HQVs and PVs are realized. However, HQVs have not been 
detected in some of NMR measurements where the system under a magnetic field perpendicular to an uniaxial anisotropy axis has been cooled through the superfluid transition temperature $T_c$ \cite{Autti2}. This observation in a field-cooled condition might imply that the HQV is not necessarily a stable object and 
require a revision on the Ginzburg-Landau (GL) description on the stability of the HQVs \cite{Nagamura,Tange}. 

In the present work, we numerically examine the vortex energy in the superfluid polar phase by using the GL free energy constructed in the weak-coupling limit and by incorporating both the dipole energy and the Fermi-liquid (FL) correction term. Throughout this paper, any vortices will be assumed to be extended along the polar anisotropy axis. We verify that, deep in the polar phase, a HQV-pair of which the size is of the order of $10 \mu m$ or longer is 
stabilized in a relatively high magnetic field. On the other hand, it is found that, in a high magnetic field applied in a tilted direction from the polar anisotropy axis, the HQV-pair does not become stable close to $T_c$ due to an enhanced role of the dipole energy. Based on these results and the presence of the vortex-pinning effect in real aerogels, we explain why no HQV has been detected in field-cooled experiments where the field direction is tilted from the polar anisotropy axis. The present result supports the previous GL theory \cite{Nagamura,Tange} concluding that the HQV should appear as a stable topological excitation even without the vortex-pinning effect due to the aerogel structure in the polar phase and other polar-distorted superfluid phases at lower temperatures. 

This paper is organized as follows. In sec.2, the model and the details of our analysis are explained, and the obtained numerical results are discussed in sec.3. In sec.4, experimental results are discussed based on the results in the preceding sections. 

\section{Model}

To study the stability of a topological excitation or a texture of a hydrodynamical variable of the superfluid order parameter, the GL free energy is a natural starting model. In the weak-coupling approximation valid only at lower pressures, the coefficients of the various terms in the GL free energy are easily evaluated using a simple model of the impurity scattering due to the aerogel structure \cite{Tange}. In a situation with an uniaxial anisotropy which is a reflection of the dimensionality of the aerogel structure, the resulting GL free energy has additional terms compared with the ordinary expression for the isotropic bulk liquid and is given, as the sum of the condensation energy term $f_{cond}$ and the quadratic gradient energy term $f_{grad}^{(2)}$, in the form 
\begin{equation}
f_{\rm GL} = f_{cond} + f_{grad}^{(2)}, 
\end{equation}
where 
\begin{eqnarray}
f_{cond}\!\!&=&\!\!(\alpha+(\alpha_z-\alpha)\delta_{iz})A_{\mu i}A^*_{\mu i}+\beta_z |A_{\mu z}A_{\mu z}^*|^2 \nonumber \\ 
&+&\beta_1^{(0)}|A_{\mu i}A_{\mu i}|^2 +\beta_2^{(0)}(A_{\mu i}A_{\mu i}^*)^2 
+ \beta_3^{(0)} A_{\mu i}^*A_{\nu i}^*A_{\mu j}A_{\nu j} \nonumber \\
&+& \beta_4^{(0)} A_{\mu i}^*A_{\nu i}A_{\nu j}^*A_{\mu j} + \beta_5^{(0)} A_{\mu i}^*A_{\nu i}A_{\nu j}A_{\mu j}^* \nonumber \\
&+& [\beta_{1}^{(1)}A_{\mu i}A_{\mu i}A_{\nu z}^*A_{\nu z}^* + \beta_{2}^{(1)} A_{\mu i}A_{\mu i}^*A_{\nu z}A_{\nu z}^* \nonumber \\
&+& \beta_{3}^{(1)} A_{\mu i}^*A_{\nu i}^*A_{\mu z}A_{\nu z} +\beta_{4}^{(1)} A_{\mu i}^*A_{\nu i}A_{\nu z}^*A_{\mu z} \nonumber \\
&+& \beta_{5}^{(1)} A_{\mu i}^*A_{\nu i}A_{\nu z}A_{\mu z}^* +c.c.], \label{eq:23}
\end{eqnarray}
and 
\begin{eqnarray}
f_{grad}^{(2)} &=& 2K_1\partial_i A_{\mu i}\partial_j A_{\mu j}^* +K_2\partial_i A_{\mu j}\partial_i A_{\mu j}^*
+K_3\partial_z A_{\mu i}\partial_z A_{\mu i}^* \nonumber\\
&+&  K_4\partial_i A_{\mu z}\partial_i A_{\mu z}^*
+K_5(\partial_i A_{\mu i}\partial_z A_{\mu z}^* + \rm{c.c.}) \nonumber \\
&+& K_6 \partial_z A_{\mu z}\partial_z A_{\mu z}^*, \label{eq:24}
\end{eqnarray}
where $A_{\mu,j}$ is the order parameter of the $p$-wave superfluid with its amplitude $|\Delta|$
. Derivation of the coefficients in eqs.(\ref{eq:23}) and (\ref{eq:24}) have been presented elsewhere \cite{Tange}. Here, the expressions of $\alpha$ and $\alpha_z$ ($\leq \alpha$), which are the most important coefficients among them, will be given in Appendix

One needs to incorporate a couple of other energy terms in the conventional GL terms given above to examine the stability of a HQV in the situation of the real NMR experiments \cite{Autti1,Autti2}. One is the FL correction energy \cite{Leggett,RS} which is one part of O($|\Delta|^4$) gradient energy terms and is directly associated with the stability of the HQV \cite{Nagamura}, where $|\Delta|$ expresses the averaged amplitude of the order parameter $A_{\mu,j}$. This energy term is given 
by 
\begin{eqnarray}
F_{\rm {FLgrad4}} &\simeq& 2 N(0) \biggl(\frac{7 \zeta(3)}{60 \pi T} \biggr)^2 \biggl(\frac{v_{\rm F}}{2 \pi T} \biggr)^2 \Gamma_1^{(s)} \nonumber \\
&\times& \int_{\bf r} {\rm Im}f_{\mu \mu,l}
\cdot {\rm Im}f_{\lambda \lambda,l}, 
\label{FL}
\end{eqnarray}
where $F_1^{(s)}=3\Gamma_1^{(s)}/(3 - \Gamma_1^{(s)})$ is the $l=1$ component of the spin-symmetric Landau parameter, $N(0)$ is the density of states on the Fermi surface, $v_{\rm F}$ is the Fermi velocity, $T$ is the temperature, and 
\begin{equation}
f_{\mu \nu, l} = A^*_{\mu s} \partial_l A_{\nu s} + A^*_{\mu l} \partial_j A_{\nu j} + A^*_{\mu j} \partial_j A_{\nu l}. 
\label{eq:fmunu}
\end{equation}
This term is necessary to make the GL approach consistent with the corresponding analysis in the London limit \cite{SV}. We note that, in the GL expansion in $|\Delta|$, the FL correction energy is one part of the O($|\Delta|^4$) gradient energy terms, and that all the O($|\Delta|^4$) gradient energy terms have to be taken into account in the numerical calculations. 
Such a whole expression of the O($|\Delta|^4$) terms is lengthy and will not be given here \cite{com}. 

In our previous studies, we have shown the stability of the HQV in the polar phase and other superfluid phases at lower temperatures to be realized in one-dimensional anisotropic aerogels by assuming the situation in zero field or in a magnetic field parallel to the polar anisotropy axis (${\bf H} \parallel {\hat z}$). Hereafter, the polar anisotropy axis characterizing the one-dimensional aerogel's structure will be identified with the $z$-axis. In the latter situation, according to eqs.(\ref{magLondon}) and (\ref{dipoleLondon}) to be given later, the configuration ${\hat {\bf d}} \perp {\bf H}$ is kept without any cost of the dipole energy. Therefore, the dipole energy was unnecessary to examine the two situations mentioned above. On the other hand, for the present purpose of examining the stability of the HQV against a {\it perpendicular} magnetic field (${\bf H} \perp {\hat z}$), we need to 
incorporate the magnetic energy and the dipole energy further. If, as usual, assuming the particle-hole asymmetry to be small enough, a $H$-linear term in 
the free energy inducing the A$_1$-phase is negligible, and we have only to focus on O($H^2$)-corrections to the GL free energy. Up to the quadratic order in the order parameter, the magnetic energy density takes the form \cite{VW} 
\begin{equation}
f_{H} = N(0) \frac{7}{12} \frac{\zeta(3)}{1+F_0^{(a)}} \biggl(\frac{\gamma \hbar}{2 \pi T} \biggr)^2 \, H_\mu H_\nu A_{\mu,j}^* A_{\nu,j}, 
\label{magf}
\end{equation}
where $\gamma \hbar/2$ is the nuclear magnetic moment of the $^3$He atom, $F_0^{(a)}$ is the $l=0$ component of the spin-antisymmetric Landau parameter, and, for simplicity, impurity scattering effects on $f_H$ were assumed to be negligibly small as well as those on $f_{\rm {FLgrad4}}$ \cite{Tange}. 
In the pure polar phase with $A_{\mu,j} = \Delta {\hat d}_\mu {\hat m}_j$, eq.(\ref{magf}) becomes 
\begin{equation}
f_H^{\rm {pol}} = g_m N(0) |\Delta|^2 ({\hat d}\cdot{\bf H})^2, 
\label{magLondon}
\end{equation}
where $g_m N(0)$ is the coefficient of eq.(\ref{magf}). According to sec.7 in Ref.\cite{VW}, $g_m$ is estimated to be $5.0 \times 10^{-8}$ [(${\rm mT})^{-2}$] at zero pressure ($P=0$(bar)). 

On the other hand, it is well known \cite{VW} that the dipole energy density takes the form 
\begin{equation}
f_{D} = \frac{3}{5} \lambda_D N(0) \biggl[ A_{j,j}^* A_{l,l} + A_{j,l}^* A_{l,j} - \frac{2}{3} A_{\mu,j}^* A_{\mu,j} \biggr], 
\end{equation}
which, in the pure polar phase, becomes 
\begin{equation}
f_D^{\rm {pol}} = \frac{6}{5} \lambda_D N(0) |\Delta|^2 \biggl[ ({\hat d}\cdot{\hat m})^2 - \frac{1}{3} \biggr]. 
\label{dipoleLondon}
\end{equation}
Here, the dimensionless coefficient $\lambda_D$ is the dipole interaction strength between the normal quasiparticles divided by the Fermi energy and is estimated to be $1.0 \times 10^{-6}$ \cite{VW}. 
Then, in a strong field situation defined by $H > H_{th} \simeq 4$ (mT), the ${\hat {\bf d}}$-vector predominantly takes the planar configuration ${\hat {\bf d}} \perp {\bf H}$, and the dipole energy may be regarded as a perturbation. The NMR experiments in which the HQV was detected \cite{Autti1,Autti2} have been performed in $H=12$ (mT). Therefore, when such a magnetic field is applied parallel to the $z$-axis, the ${\hat {\bf d}}$-vector is confined to the $x$-$y$ plane with no cost of the dipole energy. On the other hand, when a larger magnetic field than $H_{th}$ is applied along the $x$-axis, the ${\hat {\bf d}}$-vector can be assumed to be confined to the $y$-$z$ plane from the outset at least at shorter length scales than the dipole coherence length $\xi_D$ \cite{VW}. 

\section{Numerical Results}

In this section, we explain numerical results following from the extended GL free energy defined in the preceding section. 

Our numerical analysis will be performed in the same manner as in Refs.\cite{Nagamura,Tange}. As a method of solving the variational GL equations following from the GL free energy functional defined in sec.2, the direct two-dimensional method \cite{Thuneberg} is adopted by choosing the representation in the London limit of the HQV-pair configuration with the size $2a$ as the initial configuration for each numerical run. For instance, in the case where both the ${\hat d}$-vector and 
a HQV-pair with the size $2a$ lie in the $x$-$y$ plane, the order parameter $A_{\mu,j}$ in the London limit is expressed by 
\begin{equation}
A_{\mu,j} = \frac{|\Delta|}{2} [ ({\hat x}-{\rm i}{\hat y})_\mu e^{{\rm i}\phi_+} + ({\hat x}+{\rm i}{\hat y})_\mu e^{{\rm i}\phi_-} ] {\hat z}_j, 
\label{Hz}
\end{equation}
where $\phi_\pm = {\rm tan}^{-1}[y/(x \mp a)]$. 
In all numerical runs performed previously according to this method, any change of the size of the HQV-pair occurring whenever solving the GL equations was negligibly small. Consequently, the initial configuration of the order parameter expressing a HQV-pair has also played the role of the outer boundary condition for the texture of the order parameter. In fact, if eq.(\ref{Hz}) was used as the intial condition, we find that, even in the final solution, the components $A_{z,j}$ of the order parameter vanish everywhere including the close vicinity of the vortex cores. In addition, even the components, $A_{\mu,x}$ and $A_{\mu,y}$, do not appear in the final solution of a HQV-pair, possibly as a result of the strong one-dimensional anisotropy. Consequently, as have been verified in the previous work \cite{Nagamura}, a HQV-pair in he polar phase is expressed by just two components $A_{x,z}$ and $A_{y,z}$ of the order parameter and hence, is conveniently well described in the London limit. 

All the numerical computations have been performed at this time using the impurity scattering strength $(2 \pi \tau)^{-1}=0.118$ (mK), the anisotropy $\delta_u=4.4$, the pressure ($P$) dependence \cite{VW} of the superfluid transition temperature $T_{c0}$ of the bulk liquid, and the system sizes $50$($\mu$m) along the $x$-axis and $10$($\mu$m) along the $y$-axis. The definition of $\tau$ and $\delta_u$ was given in Appendix. 

\begin{figure}[t]
\begin{center}
{
\includegraphics[scale = 0.5]{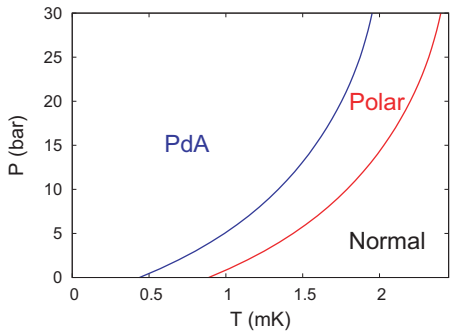}
}
\caption{$P$-$T$ phase diagram obtained according to the treatment explained in the text. At $P=3$(bar), the normal to polar transition temperature $T_c$ is $1.251$(mK), and the polar to PdA transition occurs at $T=0.8466$(mK). We note that the resulting $T_c(P)$ curve is quite close to the bulk line $T_{c0}(P)$ even for the intermediate value $\delta=4.4$ of the anisotropy (see Appendix regarding the definition of $\delta$) \cite{Hisamitsu}. 
}
\label{s:fig:VC}
\end{center}
\end{figure}

To determine the $P$ and $T$ values for which the vortex solution is examined, we need to construct the $P$ v.s. temperature ($T$) phase diagram. The normal to polar transition line $T_c(P)$ is defined as the line on which $\alpha_z=0$, while the polar to PdA continuous transition line \cite{AI06} $T_{pA}(P)$ is determined as the line on which the O($\delta \Delta^2$) term of the GL free energy vanishes, where $\delta \Delta$ is the transverse component of the PdA order parameter $A_{\mu,j} = {\hat d}_\mu(\Delta {\hat z}_j + {\rm i} \delta \Delta {\hat x}_j)$ just below the polar to PdA transition. 
One example of the resulting phase diagrams is presented in Fig.1. Hereafter, based on Fig.1, the pressure value is fixed to be $3$ (bar) at which the weak-coupling approximation will be justified. Further, we will focus on studying the vortex solutions just below $T_c(P=3({\rm bar}))=1.251$ (mK) and just 
above $T_{pA}(P=3({\rm bar})) = 0.8466$ (mK). 

\begin{figure}[tbp]
\begin{center}
{\includegraphics[scale = 0.5]{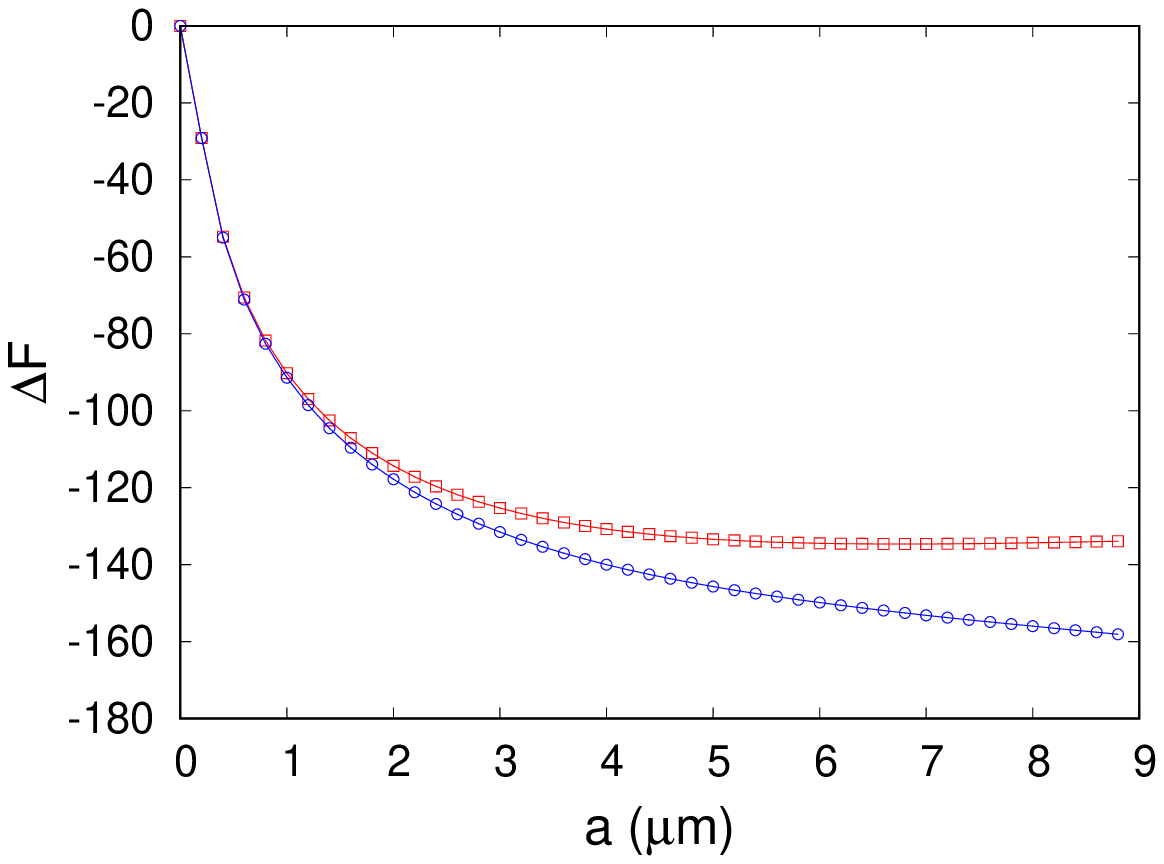}}
{\includegraphics[scale = 0.5]{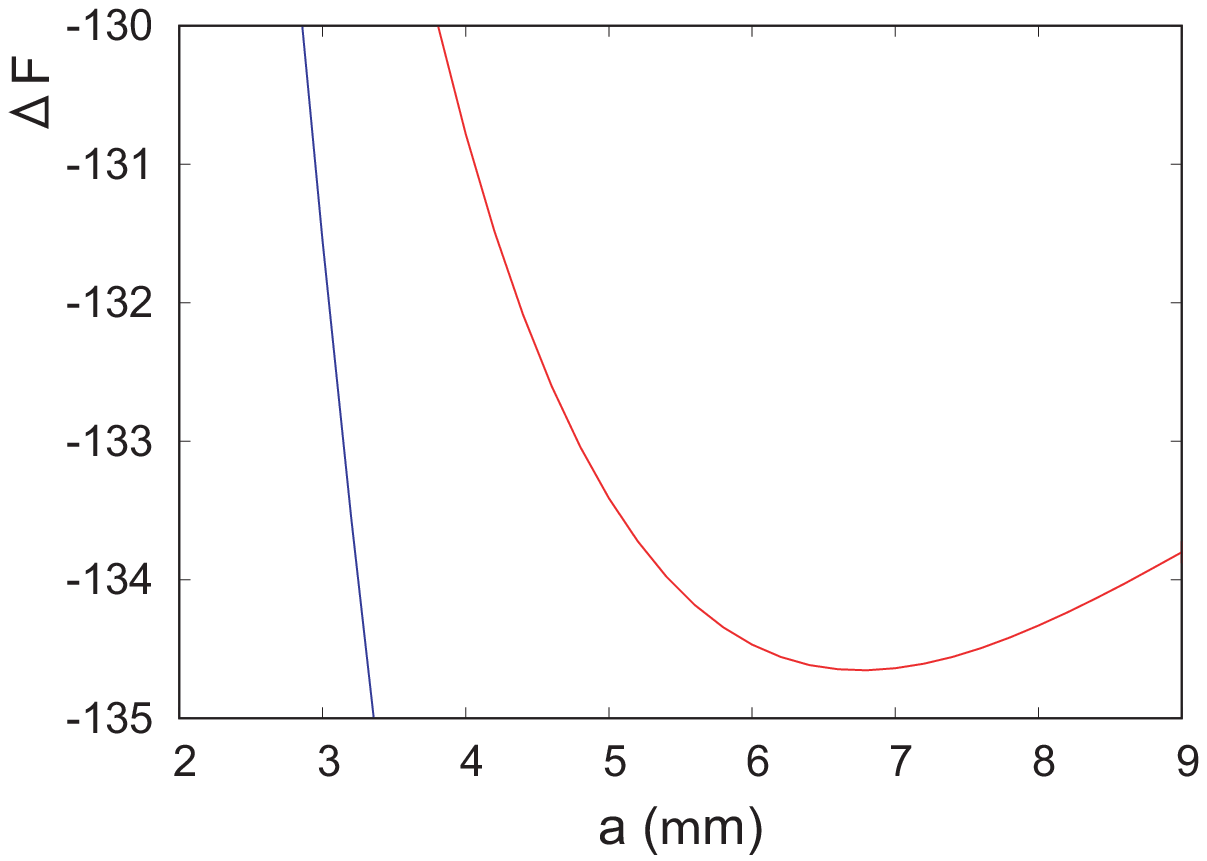}}

\caption{(a) $\Delta F=F(a)-F(0)$ v.s. $a$ (one half of the HQV-pair size) curves in an applied field parallel to the $z$-axis (blue curve and open circles) 
and perpendicular to the $z$-axis (red curve and open squares) at $T=0.85$(mK). (b) Extended view of (a) which clarifies the presence of a minimum at $a=6.8$($\mu m$) in the red curve. 
}
\label{s:fig:angle}
\end{center}
\end{figure}

\begin{figure}[tbp]
\begin{center}
{
\includegraphics[scale = 0.5]{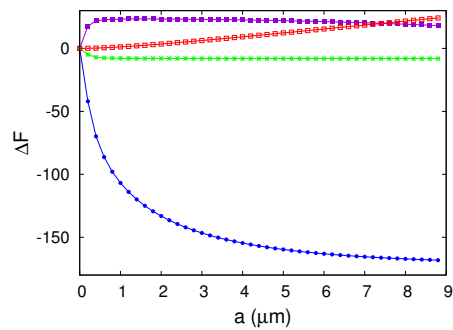}
}
\caption{Contributions of $f_{cond}$ (green curve and asterisks), $f_{grad}^{(2)}$ (purple curve and solid squares), $f_{D}$ (red curve and open squares), and the total O($|\Delta|^4$) gradient energy density including $f_{\rm {FLgrad4}}$ (blue curve and solid circles) to the red curve in the upper figure of Fig.2. The expression of the total O($|\Delta|^4$) gradient energy density is given in (A17) of Ref.6. 
}
\label{s:fig:angles}
\end{center}
\end{figure}

First, let us discuss about results on a HQV-pair appearing at $T=0.85$(mK) close to $T_{pA}$. As mentioned above, the structure of a HQV-pair in the polar phase far below $T_c$ is expressed by just two components of $A_{\mu,j}$ and thus, is well described in the London limit \cite{Nagamura}. The blue curve consisting of the open circle symbols is the curve of the vortex energy obtained in terms of the initial configuration (\ref{Hz}). As explained below eq.(\ref{Hz}), this corresponds to the case under a magnetic field parallel to the $z$-axis or zero field. Quite a similar result has been shown in our previous work \cite{Nagamura}. In Fig.2, $\Delta F$ is the energy difference $F(a) - F(0)$, where $F(a)$ is the energy of a HQV-pair with the size $2a$. Thus, the negative $\Delta F$ implies the stability of a HQV-pair against a PV in ${\bf H} \parallel {\hat z}$ or in zero field. 

On the other hand, the red curve consisting of the open square symbols in Fig.2 is essentially a new result. In this case where, as the initial condition, the configuration 
\begin{equation}
A_{\mu,j} = \frac{|\Delta|}{2} [ ({\hat z}+{\rm i}{\hat y})_\mu e^{{\rm i}\phi_+} + ({\hat z}-{\rm i}{\hat y})_\mu e^{{\rm i}\phi_-} ] {\hat z}_j, 
\label{Hx}
\end{equation}
was used to represent the situation in which the magnetic field is parallel to the $x$-axis, and the incorporated dipole energy becomes effective. 
At longer distances, the dipole energy dominates over the FL-correction term (\ref{FL}), lowering the energy of a HQV-pair with increasing $a$. Consequently, as is seen in the lower figure of Fig.2, $\Delta F$ gets a minimum at $a=6.8$($\mu$m), implying that the resulting size of the HQV-pair becomes $13.6$($\mu$m) comparable with the dipole coherence length $\xi_D$. 

Figure 3 expresses the resulting contributions of various energy terms to $\Delta F$. Clearly, the primarily negative $\Delta F$-value is dominated by the FL-correction term (\ref{FL}). The feature that the quadratic gradient term shows a sharp growth only in the range $a \leq 0.2$($\mu$m) suggests that the core radius of a single PV is at most $0.2$($\mu$m). The dipole energy induces a deviation from the planar configuration ${\hat {\bf d}} \perp {\bf H}$ of the ${\hat {\bf d}}$-vector, and consequently, the contribution of the dipole energy to $\Delta F$ monotonously increases with increasing $a$ defined in the initial configuration (\ref{Hx}). 

Therefore, deep in the polar phase, a HQV-pair with the size of the order of $\xi_D$ is stabilized even in a strong field perpendicular to the polar anisotropy axis \cite{SV}. 

\begin{figure}[tbp]
\begin{center}
{\includegraphics[scale = 0.5]{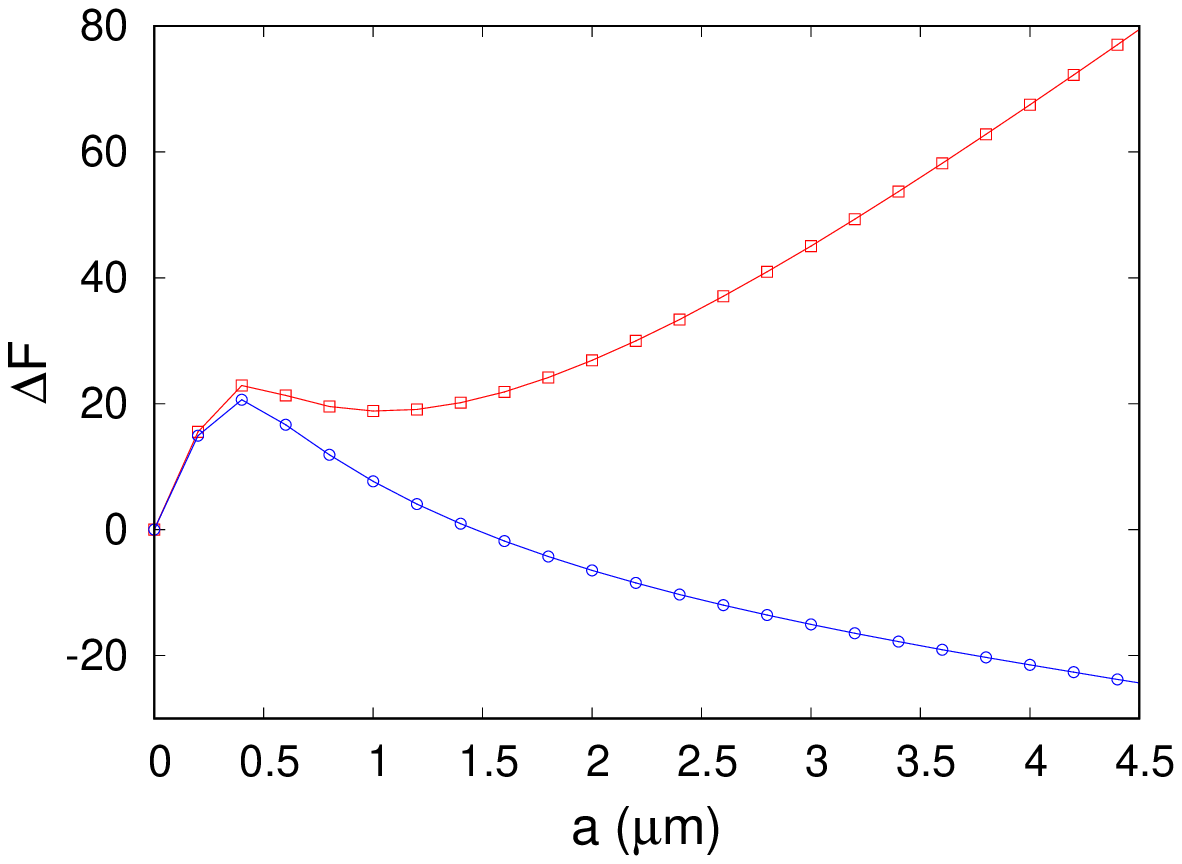}}
{\includegraphics[scale = 0.5]{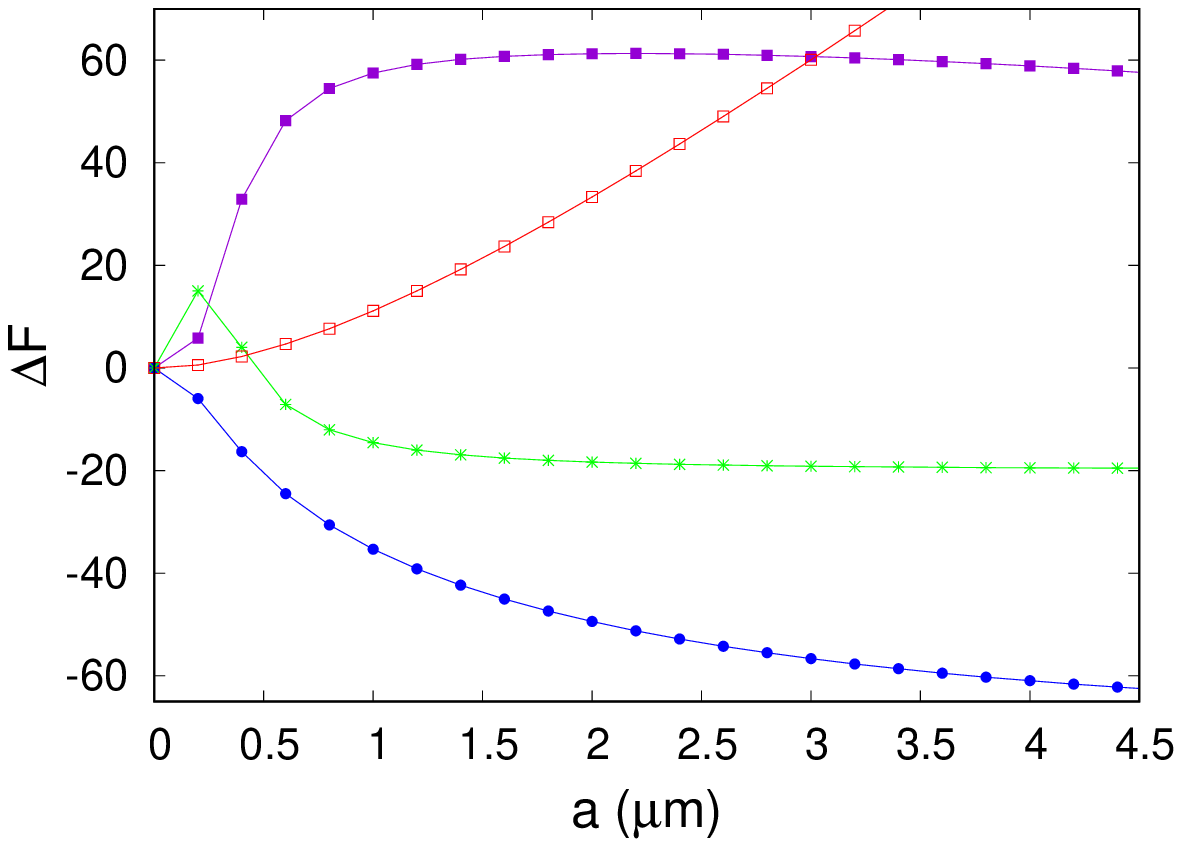}}
\caption{(Upper figure) Corresponding results to the upper figure of Fig.2 at $T=1.22$(mK) just below $T_c=1.251$(mK). (Lower figure) Contributions of $f_{cond}$ (green curve and asterisks), $f_{grad}^{(2)}$ (purple curve and solid squares), $f_{D}$ (red curve and open squares), and the total O($|\Delta|^4$) gradient energy density (blue curve and solid circles) to the red curve in the upper figure. 
}
\label{s:fig:angle4}
\end{center}
\end{figure}

\begin{figure}[tbp]
\begin{center}
{\includegraphics[scale = 0.35]{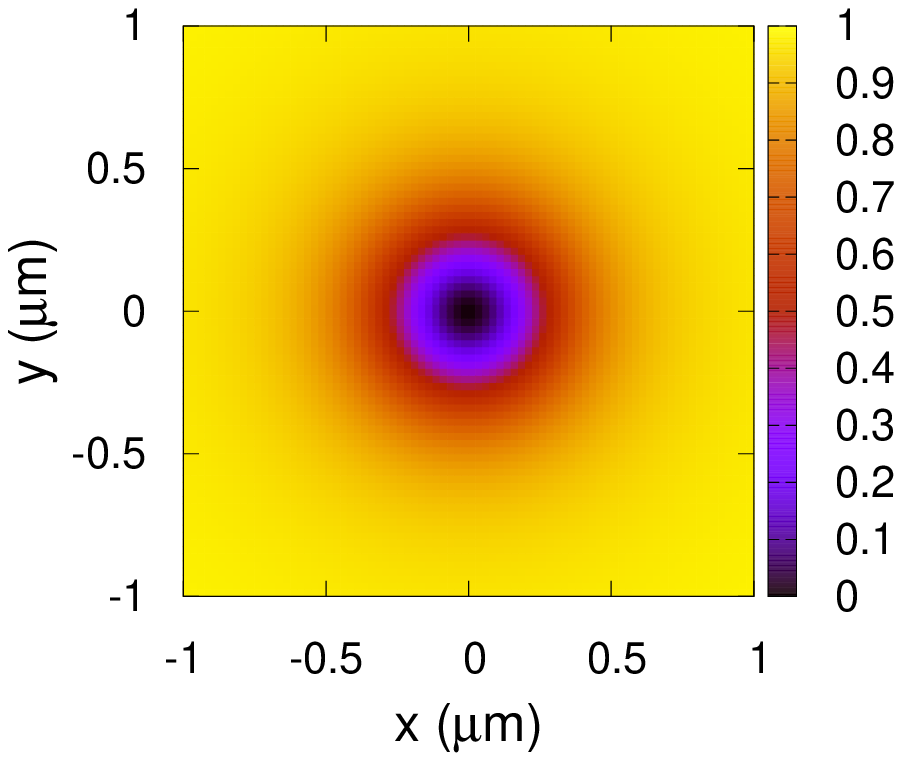}}
{\includegraphics[scale = 0.35]{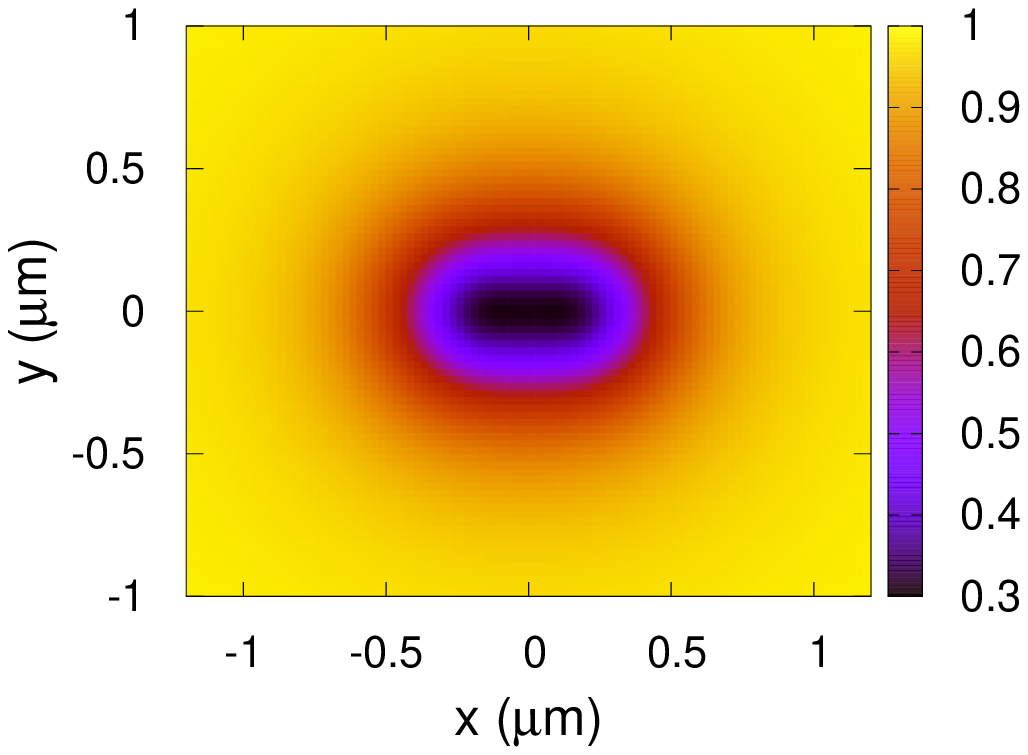}}
{\includegraphics[scale = 0.35]{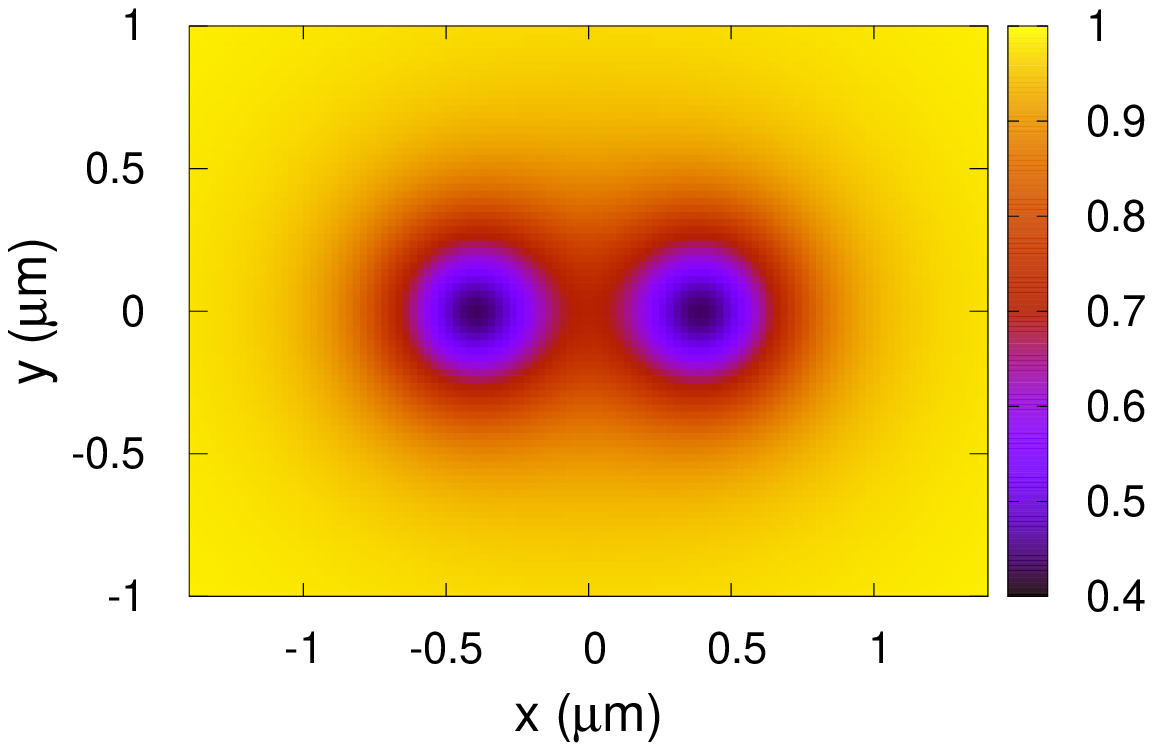}}
\caption{Spatial profiles of the squared amplitude $\sum_{\mu,j} |A_{\mu,j}|^2$ in the $x$-$y$ plane at $a=0$ (top), $a=0.2$($\mu$m) (middle), and $a=0.4$($\mu$m) (bottom) in the case of the red curve in the upper figure of Fig.4. 
}
\label{s:fig:angle5}
\end{center}
\end{figure}

Next, we turn to the corresponding results at $T=1.22$(mK) just below $T_c$ which are presented in Fig.4. If the contribution of the dipole energy is negligibly small, or the applied magnetic field is parallel to the polar anisotropy axis ($z$-axis), the quadratic gradient energy becomes comparable in magnitude with the FL-correction at least for lower $a$-values. In a situation where the strong coupling correction is quantitatively important, $\Delta F$ remains positive so that a PV rather than a HQV-pair is created \cite{Nagamura}. In the present weak-coupling approximation, however, the FL-correction term dominates over the quadratic gradient term for larger $a$-values, and, as the blue curve composed of the open circle symbols shows in the upper figure of Fig.4, a HQV-pair with a large size tends to be created. Therefore, on cooling through $T_c$ in zero field or in a field parallel to the polar anisotropy axis, the HQV-pair should be created even near $T_c$ at such a low pressure that the weak-coupling approach is justified. 

It should be noted that the $\Delta F$ v.s. $a$ curve has shown a local maximum close to the $a=0.4$($\mu$m)-value. Such a maximum in the $\Delta F$ v.s. $a$ curve has also been found in a previous analysis \cite{Nagamura} on the vortices in the polar phase occurring under a weak anisotropy. There, both the FL-correction and the strong-coupling correction necessary for the chiral phase in the bulk liquid were incorporated in the GL free energy, and the resulting maximum at a low $a$-value appearing even at a low pressure has been ascribed to a weak effect of the strong-coupling correction. The presence of this maximum in the present weak-coupling approach implies that this interpretation given in Ref.\cite{Nagamura} has to be revised. The presence of this local maximum becomes important in understanding the experimental observation \cite{Autti1,Autti2} that no HQV have appeared in the perpendicular field configuration (see below). 

On the other hand, once the dipole energy is incorporated, in a magnetic field perpendicular to the $z$-axis, it brings an energy cost for the HQV-pair. In this case, as the lower figure of Fig.4 shows, the FL-correction energy decreasing with $a$ is overcome in $\Delta F$ by the dipole energy sharply increasing with $a$, reflecting the fact that the former is quartic in the order parameter and thus, of a higher order in the GL expansion compared with the latter. Consequently, the total $\Delta F$-curve close to $T_c$, the red curve in the upper figure, remains positive and increases with $a$ so that just a PV becomes stable in a strong field perpendicular to the $z$-axis. It will be clear that the presence of the local maximum, already mentioned in ${\bf H} \parallel {\hat z}$ case, plays a key role in obtaining the $\Delta F(a)$ curve keeping positive for any $a$-value. Without such a local maximum, a HQV-pair with a small but finite size would appear even near $T_c$. 

Interestingly, as the lower figure of Fig.4 shows, the direct origin of the local maximum close to $a=0.4$($\mu$m) is a rapid {\it increase} of the condensation energy $f_{cond}$ in a range of small $a$ values. At the maximum where the PV tries to be splitted to two HQVs, there is an energy barrier against the separation. In fact, the extension of the PV's core (see the middle figure of Fig.5) seems to lead to a large energy cost due to a variation of the amplitude of the order parameter close to $T_c$. Far below $T_c$ where the London limit is the safely valid description, such an $a$-dependent change of the condensation energy is not seen (see Fig.2). In Ref.\cite{Nagamura} where the numerical analysis has been performed by always incorporating the strong-coupling correction, this unusual behavior of $f_{cond}$ near $T_c$ has been overlooked.

\section{Discussion}

The results in the preceding section imply that, when the liquid $^3$He under rotation is cooled through $T_c$ under a magnetic field perpendicular to the polar anisotropy axis and thus, to the rotation axis, a PV is created as the stable vortex, while the PV is smoothly splitted to two HQVs upon a further cooling in the same system. 

Now, it is not difficult based on this observation to explain why the HQVs have not been detected depending on the field orientation in real aerogels. As found in Ref.\cite{KLS}, large HQV-pairs created in the high temperature polar phase survive in the Polar-distorted B (PdB) phase lying at lower temperatures. Reflecting the fact that a single HQV is topologically prohibited in the PdB phase, a large HQV-pair in the polar phase {\it naturally} has to shrink to a small pair \cite{Tange}. Based on this finding, it was clarified \cite{KLS} that the large HQV-pair are seen in the PdB phase in real systems because the HQVs created in the polar phase are strongly pinned by the strands composing the aerogel which are nearly aligned along the $z$-axis. Quite a similar argument holds in the present case: When a rotated system under a magnetic field perpendicular to the $z$-axis is cooled through $T_c$, the stable vortex in the polar phase just below $T_c$ is not a HQV-pair but a PV. Although this PV should be naturally broken up into two HQVs, it is stabilized even at lower temperatures as a consequence of the strong pinning due to the aerogel strands aligned along the $z$-axis. On the other hand, under a magnetic field applied along the $z$-axis, only HQV-pairs are stabilized and pinned by the aerogel. By repeating similar measurements while changing the direction of the applied field continuously, it may be valuable to examine how the coexistence of the PVs and the HQVs occurs.

\section{Appendix} 

Here, we briefly describe how the GL free energy functional is derived based on a model of the impurity scattering via the nematic aerogel. To perform this, 
the Hamiltonian 
\begin{equation}
{\cal H}_{\rm imp} = \int_{\bf r}  u({\bf r}) {\hat n}({\bf r}). 
\label{eq:11}
\end{equation}
needs to be added to the familiar BCS Haniltonian \cite{VW}, where ${\hat n}({\bf r})$ is the particle density operator, and the scattering potential $u({\bf r})$ is assumed to have zero mean and to obey the following correlator  
\begin{eqnarray}
W({\bf r}) &=& 2 \pi N(0) \tau \, {\overline {u({\bf r}) u(0)}} 
= \frac{k_{\rm F}}{2} \, \delta^{(2)}({\bf r}_\perp) \nonumber \\ 
&\times& e^{-|z|/L_z}. 
\label{eq:15}
\end{eqnarray}
Here, $|\delta_u|=k_{\rm F}^2 L_z^2$ is the measure of the anisotropy, and $L_z$ is the correlation length on the scattering event. Then, the Fourier transform of $W({\bf r})$ becomes 
\begin{equation}
w({\bf k}) = \frac{\sqrt{|\delta_u|}}{ 1 + L_z^2 k_z^2},
\label{eq:14}
\end{equation}
Then, the fermion Green's function ${\cal G}_{\bf p}(\varepsilon)$ with the Matsubara frequency $\varepsilon$ is given in the normal state by $({\rm i}|{\tilde \varepsilon}_p|{\rm sgn}(\varepsilon) - \xi_{\bf p})^{-1}$, where 
\begin{eqnarray}
|{\tilde \varepsilon}_p| &=& |\varepsilon| + \frac{1}{2 \tau} \langle w(p - p') \rangle_{p'} \nonumber \\ 
&=& |\varepsilon| + \frac{1}{4 \tau} [ \, {\rm tan}^{-1}(|\delta_u|^{1/2}(1 -  p)) \nonumber \\
&+& {\rm tan}^{-1}(|\delta_u|^{1/2}(1 + p)) \, ] \label{eq:A1}
\end{eqnarray}
($|p| < 1$), where $p={\bf p}\cdot{\hat z}/p_{\rm F}$, $\langle \,\,\, \rangle_{p}$ implies the average over the polar angle ${\rm cos}^{-1}(p)$. 

Further, the $p$-wave pairing function ${\bf p}_i$ is renormalized through the impurity scattering by ${\bf p}_j(\delta_{ij} - {\hat z}_i {\hat z}_j)+C_0 p_z {\hat z}_i$, where 
\begin{eqnarray}
C_0 &=& \frac{1}{1 - 2(I_{d11}-I_{d12})}, \nonumber \\
I_{d1n} &=& \frac{\sqrt{|\delta_u|}}{2 \tau} \biggl\langle \frac{1}{|{\tilde \varepsilon}_p| \, (1 + |\delta_u| p^2)^n} \biggr\rangle_p. 
\label{eq:A6}
\end{eqnarray}
For instance, the coefficients of the quadratic mass terms in the GL free energy functional are given by
\begin{eqnarray}
\alpha &=& \frac{1}{3} N(0) \biggl[ {\rm ln}\biggl(\frac{T}{T_{c0}} \biggr) + 2 \pi T \sum_{\varepsilon > 0} \biggl( \frac{1}{|\varepsilon|} - \frac{3}{2} ( I_{10} - I_{11} ) \biggr) \biggr], \nonumber \\
\alpha_z &=& \frac{1}{3} N(0) \biggl[ {\rm ln}\biggl(\frac{T}{T_{c0}} \biggr) + 2 \pi T \sum_{\varepsilon > 0} \biggl( \frac{1}{|\varepsilon|} - 3  I_{11} C_0 \biggr) \biggr], \label{eq:A7}
\end{eqnarray}
where 
\begin{equation}
I_{1n} = \biggl\langle \frac{p^{2n}}{|{\tilde \varepsilon}_p|} \biggr\rangle_p. \label{eq:A8}
\end{equation}

In the limit of $L_z \to \infty$, $\alpha_z$ reduces to the expression in the clean limit (i.e., of the bulk liquid) $N(0) {\rm ln}(T/T_{c0})/3$, implying that the Anderson's Theorem is satisfied \cite{Hisamitsu}.


\begin{thebibliography}{99}
\bibitem{com} Equation (A17) of Ref.6 has been used in the present numerical analysis as the whole expresson of O($|\Delta|^4$) gradient energy.
\end{thebibliography}
\end{document}